# New Development in RF Pulse Compression

Sami G. Tantawi[*], SLAC, Menlo Park, CA94025, USA

*Abstract*

Several pulse compression systems have been proposed for future linear collider. Most of these systems require hundreds of kilometers of low-loss waveguide runs. To reduce the waveguide length and improve the efficiency of these systems, components for multimoding, active switches and non-reciprocal elements are being developed. In the multimoded systems a waveguide is utilized several times by sending different signals over different modes. The multimoded components needed for these systems have to be able to handle hundreds of megawatts of rf power at the X-band frequency and above. Consequently, most of these components are overmoded. We present the development of multimoded components required for such systems. We also present the development efforts towards overmoded active component such as switches and overmoded non-reciprocal components such as circulators and isolators.

## 1 INTRODUCTION

Rf pulse compression systems enhance the peak power capabilities of rf sources. Indeed, they have been used to match the short filling time of an accelerator structure to the long pulse length generated by most rf sources such as klystrons. All rf pulse compression system store the rf energy for a long period of time and then release it in a short time. For linac applications associated with future linear colliders, the storage medium is a waveguide transmission line. The energy required to supply a linac section or a set of linac sections is stored in these lines. The length of these waveguide transmission lines has the same order as $\tau c$ where $\tau$ is the pulse length required by the linac and $c$ is the speed of light. For colliders based on X-band linacs such as the NLC [1] and JLC [2] these lengths are tens of meters. Since the collider usually contains several-thousand accelerator sections, the total waveguide system for the collider is usually hundreds of kilometers long.

These long waveguide runs have to be extremely low-loss. At the same time they should be able to handle power levels of order hundreds of megawatts. Hence, these waveguides are usually highly overmoded circular waveguides operating under vacuum. Because of vacuum, and tolerance requirements, these hundreds of kilometers of waveguide runs are expensive, and hard to install and maintain.

To reduce these waveguide runs, several innovations have been made both on the system and component levels:

1. RF pulse compression systems that have high intrinsic efficiencies have been suggested. These systems are Binary Pulse Compression (BPC) [3], Delay Line Distribution System (DLDS) [4], and active pulse compression system using resonant delay lines [5-6].
2. Enhancing the system power handling capabilities can ultimately reduce the number of systems required. One can use a single system that serves several rf sources and several accelerator sections. Hence, low-loss overmoded components have been developed for these systems, see, for example, [7-9]
3. Since these waveguide runs are overmoded one can utilize these waveguides several times by sending signals over different modes. Such multimoded systems have been suggested [10] and conceptual tests of components and concepts have been performed [11].
4. To implement active pulse compression systems inexpensive super-high-power semiconductor switching arrays have been suggested [12], and tested [13]

In this paper we devote section 2 to an accurate formulation for the length of waveguide runs required by several pulse compression systems. We then describe in section 3 the work done to provide a super high power test setup for the components required by these systems. In section 4 we describe the multimoded planer components and associated tapers. Finally, in section 5, we show some attempts to provide a semiconductor microwave switch.

## 2 COMPARISON BETWEEN RF PULSE COMPRESSION SYSTEMS

### 2.1 General Layout

To achieve pulse compression a storage system is employed to store the rf power until it is needed. Different portions of the input rf pulse $T$ are stored for different amounts of time. The initial portion of the rf pulse is stored for a time period $t_m$, the maximum amount of storage time for any part of $T$. It is given by,

$$t_m = \tau(C_r - 1). \tag{1}$$

where $\tau$ is the accelerator structure pulse width and is given by

$$\tau = \frac{T}{C_r} \tag{2}$$

---

and $C_r$ is the compression ratio. The realization of the storage system is usually achieved using low-loss waveguide delay lines. These lines are usually guides that propagate the rf signal at nearly the speed of light. The *maximum* length required for these guides, per compression system, is

$$l^{max} = t_m v_g \frac{C_r}{2}, \quad (3)$$

where $v_g$ is the group velocity of the wave in the delay line. The total number of rf pulse compression systems required for the accelerator system is given by

$$N_c = \frac{N_a P_a}{P_k n_k C_r \eta_c}; \quad (4)$$

where $N_a$ is the total number of accelerator structures in the linac, $P_k$ is the klystron (or the rf power source) peak power, $P_a$ is the accelerator structure required peak power, $n_k$ is the number of klystrons combined in one pulse compression system, and $\eta_c$ is the efficiency of the pulse compression system.

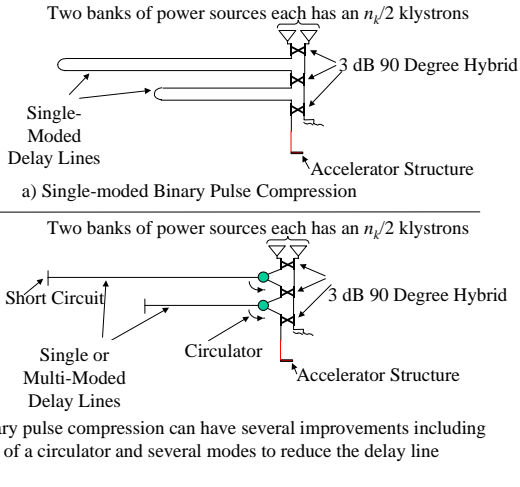

a) Single-moded Binary Pulse Compression

b) Binary pulse compression can have several improvements including the use of a circulator and several modes to reduce the delay line length.

Fig. 1 Binary Pulse Compression system

Thus the maximum total length of waveguide storage line for the entire linac is given by

$$L^{max} = l^{max} N_c = \frac{1}{n_k \eta_c} \frac{N_a P_a}{P_k} \frac{(C_r - 1)}{2} \tau v_g. \quad (5)$$

In general the total length $L$ is given by

$$L = L^{max} R_l; \quad (6)$$

where $R_l$ is a length reduction factor which varies from one system to another and, in general, is a function of the compression ratio. Finally, the total number of klystrons in the system $N_k$ is given by,

$$N_k = \frac{1}{C_r \eta_c} \frac{N_a P_a}{P_k}. \quad (7)$$

### 2.2 Binary Pulse Compression System

For details of the original single moded system the reader is referred to [3]. The system is shown in Fig. 1. The single moded BPC, in its original form, has a length reduction factor $R_l$ of $2/C_r$. It becomes more economical at higher compression ratios. However, the power being handled by the waveguides and rf components is doubled at every stage of the BPC system. Naturally, the peak power depends on the number of klystrons that one might use in one system, i.e., $n_k$. The length reduction factor is given by

$$R_l = \frac{2-c}{n_m C_r}; \quad (8)$$

where $n_m$ is the number of modes used in the system. The parameter $c$ determines the length reduction if a circulator is used and is 1 if a circulator is used and 0 otherwise.

The efficiency of the system is given by

$$\eta_c = \eta_{cir} \eta_{com} \frac{1 - 10^{-\left(\sum_{i=1}^{n_m} \frac{\alpha_i \tau}{10 n_m}\right) C_r}}{C_r \left(1 - 10^{-\left(\sum_{i=1}^{n_m} \frac{\alpha_i \tau}{10 n_m}\right)}\right)}; \quad (9)$$

where $\alpha_i$ is the attenuation constant in dB/unit time for mode $i$, and $\eta_{cir}$ and $\eta_{com}$ are the circulator efficiency and component efficiency respectively.

### 2.3 Delay Line Distribution System (DLDS)

The original description of the DLDS is found in [4]. A modification to that system with multimoded delay lines is discussed in [14]. However, accurate accounts for the efficiency and waveguide length are introduced here. The system is shown in Fig. 2. To give an expression for the length reduction factor in terms of the number of modes $n_m$ we first define the number of pipes per unit rf system as

$$n_p = \left[\frac{C_r - 1}{n_m} + 0.5\right]; \quad (10)$$

where [.] means the integer-value function. The length reduction factor is, then, given by

$$R_l = \frac{n_p \left(C_r - 1 - (n_m/2)(n_p - 1)\right)}{C_r (C_r - 1)} \quad (11)$$

The efficiency of the system is given by:

$$\eta=\frac{\eta_{on}}{C_r}\left(1+\sum_{j=1}^{n_m}\frac{10^{-\alpha_j\frac{\tau}{20}(C_r-j)}(10^{-\alpha_j\frac{\tau}{20}(n_p-j)n_m}-1)}{10^{-\alpha_j\frac{\tau}{20}n_m}-1}+\sum_{j=1}^{C_r-1-(n_p-1)n_m}10^{-\alpha_j\frac{\tau}{20}(C_r-n_m(n_p-1)-j)}\right) \quad (12)$$

where $\alpha_j$ is the attenuation of mode $j$ in dB/unit time.

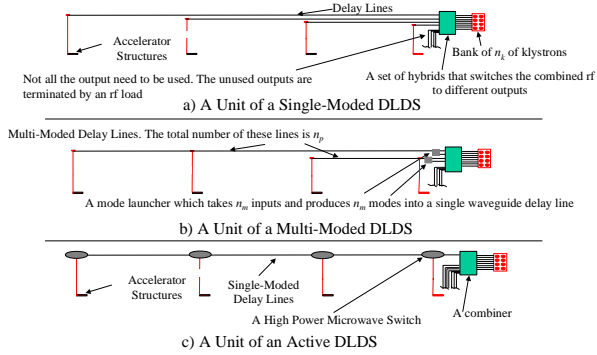

Fig. 2 Delay Line Distribution System

If a switch is used only one pipe is used and the length reduction factor becomes $1/C_r$. The efficiency in that case becomes

$$\eta=\frac{1}{C_r}\left(\frac{1-\left(\eta_s^{off}\eta_\tau\right)^{C_r-1}}{1-\eta_s^{off}\eta_\tau}\eta_s^{on}+\left(\eta_s^{off}\eta_\tau\right)^{C_r}\right); \quad (13)$$

where $\eta_s^{on}$ is the efficiency of the switch in the *on* state, while $\eta_s^{off}$ is the efficiency of the switch in the *off* state. The quantity $\eta_\tau$ is the efficiency of the waveguide due to the attenuation of that waveguide for a period of time $\tau/2$.

*2.4 Resonant Delay Lines*

The original description of the resonant delay lines can be found in [15]. An extensive analysis of the system and its variations using active switching are presented in [5]. High power experimental results and techniques are described in the next section of this article and detailed in Ref. [7].

The system and its variations are shown in Fig. 3. The length reduction factor is given by

$$R_l=\frac{2-c}{n_mC_r(C_r-1)}; \quad (14)$$

where $c$ determine the length reduction if a circulator is used and is 1 if a circulator is used and 0 otherwise. The efficiency of the system is given by

$$\eta=\frac{\eta_{cir}}{C_r}\left(R_0+\left(1-R_0^2\right)\frac{1-(R_010^{-\alpha/10\tau})^{C_r-1}}{1-R_010^{-\alpha/10\tau}}10^{-\alpha/10\tau}\right)^2; \quad (45)$$

where $\alpha$ is the attenuation /unit-time in dB and is given by $\alpha=\frac{1}{n_m}\sum_{i=1}^{n_m}\alpha_i$; and $\alpha_i$ is the attenuation/unit time for mode $i$. The parameter $R_0$ is a function of the compression ratio [5] and is, approximately, given by

$$R_0(C_r)\approx 0.871-0.514e^{-0.164C_r}, C_r\leq 24.$$

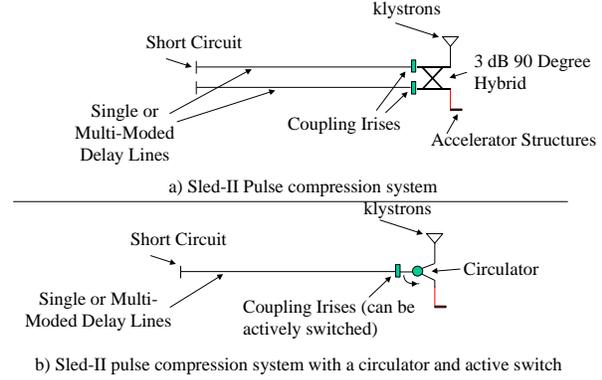

Fig. 3 Resonant Delay Line Pulse Compression System

If one can design and implement a super-high-power switch, the intrinsic efficiency of the SLED-II system can be enhanced. Intrinsic efficiency of this system is approximately 82% [5], and the total efficiency is slightly reduced from that number. The efficiency in this case has a weak dependence on the compression ratio.

*2.4 Comparison*

Table 1 shows the parameters of different single-moded pulse compression systems if used with the current design of the Next Linear Collider [1]. Clearly, these systems comprise very long runs of low-loss vacuum waveguide. Several innovations are required to reduce the length and to make these systems operate at these high power levels. These are discussed in the following sections.

| System | $C_r$ | Waveguide Length | $\eta$ (%) | Peak Power | Number Of Klystrons |
|---|---|---|---|---|---|
| DLDS | 4 | 131 km | 85 | 600 MW | 3168 |
|  | 8 | 305 km | 85 | 600 MW | 1584 |
| BPC | 4 | 523 km | 85 | 600 MW | 3168 |
|  | 8 | 698 km | 85 | 1200 MW | 1584 |
| (SLED-II) | 4 | 180 km | 82 | 493 MW | 3277 |
|  | 8 | 124 km | 59 | 716 MW | 2258 |

Table 1: Parameters of single-moded different pulse compression systems

**3 HIGH POWER IMPLEMENTATION OF THE RESONANT DELAY LINE SYSTEM (SLED-II)**

More technical details for the high power SLED-II system can be found in [7]. Here we summarize the design and the obtained results.

To separate the input signal from the reflected signal, one might use two delay lines fed by a 3-dB hybrid as shown in Fig. 4. The reflected signal from both lines can be made to add at the forth port of the hybrid. Fig.4 shows the pulse-compression system. For delay lines, two 22.48-meter long cylindrical copper waveguides are used; each is 12.065 cm in diameter and operates in the $TE_{01}$ mode.

In theory, these overmoded delay lines can form a storage cavity with a quality factor $Q > 1 \times 10^6$. A shorting plate, whose axial position is controllable to within ±4 µm by a stepper motor, terminates each of the delay lines. The input of the line is tapered down to a 4.737 cm diameter waveguide at which the $TE_{02}$ mode is cutoff; hence, the circular irises which determine the coupling to the lines do not excite higher order modes provided that they are perfectly concentric with the waveguide axis.

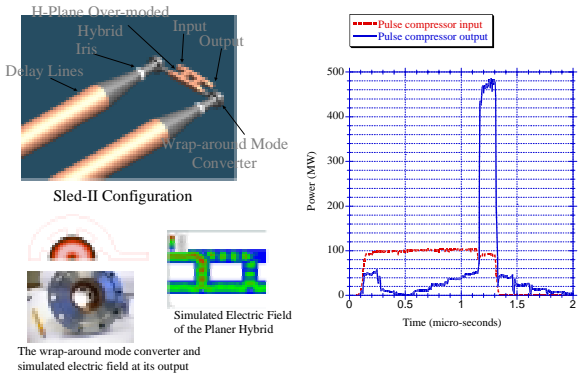

Fig. 4 The High Power SLED-II System

A compact low-loss mode converter excites the $TE_{01}$ mode just before each iris [7]. These mode transducers, known as wrap-around mode converters, were developed specifically for this application. The mode converters are connected to two uncoupled arms of a high-power, over-moded, planar 3-dB hybrid. This hybrid has been designed especially for this application so that it can handle the super high power produced by this system [9]. The distance from the irises to the center of the hybrid has been adjusted to within ±13 µm to minimize reflections to the input port. The iris reflection coefficient is optimized for a compression ratio of 8.

The system is designed to operate under vacuum. All the components are designed to handle the peak fields required by the high power operating conditions of the system. At 11.424 GHz and 600 MW peak power the maximum field level is less than 40 MV/m.

The input and output pulse shapes of the system are shown in Fig. 4. The output pulse reached levels close to 500 MW. It was limited only by the power available from the klystrons.

## 4 MULTIMODED STRUCTURES

A multimoded system was first suggested for the DLDS system [14]. Several designs for multimoded components have recently been developed [16]. However, the most promising set of components are those based on planer microwave structures [17]. These were an extension to the planner hybrid designs developed for the high-power SLED-II pulse compression system (see section 3 of this article). These planer structures have the advantage of a design that is insensitive to its height. Hence one can increase the components height to any desired value to reduce the peak rf fields at the walls.

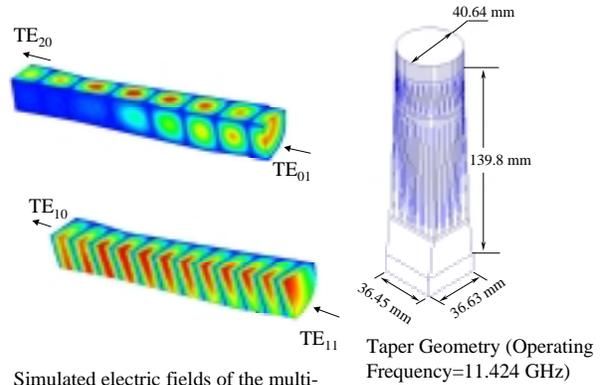

Fig. 5 Multimoded circular to rectangular taper

To transfer the rectangular waveguide cross section of these components into a circular waveguide cross section, thus making them compatible with the circular waveguide delay lines, one needs a special type of taper. Tapers that transform waveguide modes from circular to rectangular have been reported in [8]. These tapers could be extended to operate with several modes at once. An example of such a taper is shown in Fig. 5. The tapers take the input of a near square waveguide carrying the $TE_{10}$ and the $TE_{20}$ modes and transfer them into the circular waveguide modes $TE_{11}$, and $TE_{01}$ respectively. These tapers perfectly match the planar multimoded launcher and extractors presented in [17].

## 5 ACTIVE SYSTSEMS

Super-High-Power microwave switches can reduce the cost of the DLDS while increasing its capabilities for higher compression ratios. When used with DLDS one can use one single pipe as shown in Fig. 2.

**PIN diode array Active Window**
- All doping profile and metallic terminals on the window are radial, i.e. perpendicular to electric field of the $TE_{01}$ mode. → Effect of doping and metal lines on RF signal is small when the diode is reverse biased.
- With forward bias, carriers are injected into I region and I region becomes conductor → RF signal is reflected.

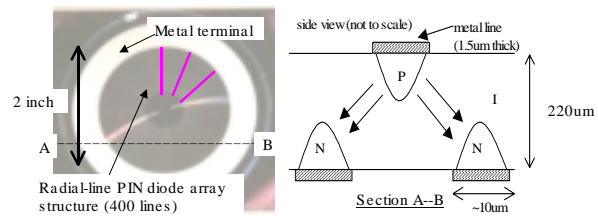

- Base material: high resistivity (pure) silicon, <5000ohm-cm, n-type
- Diameter of active region: 1.3 inch
- Thickness: **220um**
- Coverage (metal/doping line on the surface): ~10%

Fig. 6 Implementation of a PIN diode active window

With resonant delay line systems active switches can dramatically improve their efficiencies making it possible to utilize these compact systems for linear collider applications.

Indeed, these active switches have attracted the attention of numerous researchers. However, most of the concepts that were suggested are either very expensive or impractical. A promising concept which combines both economical aspects and practical designs were suggested recently [13]. The use of a several elements of such a switch was explored [12]. The switch is shown in Fig. 6. The window shown operates in a waveguide carrying the $TE_{01}$ mode. Hence all the electric field lines are normal to the doping and metalization lines. Because the $TE_{01}$ mode does not carry any axial currents the separation of the waveguide to supply the diodes with the required bias was possible. These switches operated at power levels around 10 MW at 11.424 GHz. This exceeds by orders of magnitude the capabilities of any known semiconductor rf switch.

## 6 SUMMARY

Several pulse compression systems have developed for use with the rf system of future linear colliders. These systems suffer from very long waveguide runs. Some of the systems that have a compact nature also suffer from efficiency degradation. To improve these systems several innovations were introduced. These innovations increase power-handling capabilities, make the system more compact by utilizing several modes within a single waveguide, and finally improve the system's layout and performance by turning them into active systems.

## 7 ACKNOWLEDGMENTS

This work reported herein represents the collaborative effort of several people, a partial list of them is mentioned here: C. Nantisat, N. Kroll, P. Wilson, F. Tamura, R. Ruth, G. Bowden, R. Lowoen, V. Dolgashev, K. Fant, A. Vlieks, R. Fowkes, C pearson, A. manegat, and the klystron mechanicalwork shop personel at SLAC.
This work is supported by Department of Energy contract DE–AC03–76SF00515.

## REFERENCES

[1] The NLC Design Group, Zeroth-Order Design Report for the Next Linear Collider, LBNL-PUB-5424, SLAC Report 474, and UCRL-ID 124161, May 1996

[2] The JLC Design Group, JLC Design Study, KEK-REPORT-97-1, KEK, Tsukuba, Japan, April 1997.

[3] Z.D. Farkas, "Binary Peak Power Multiplier and its Application to Linear Accelerator Design," IEEE Trans. MTT-34, 1986, pp. 1036-1043.

[4] H. Mizuno and Y. Otake, "A New Rf Power Distribution System for X Band Linac Equivalent to an Rf Pulse Compression Scheme of Factor $2^N$," 17th International Linac Conference (LINAC 94), Tsukuba, Japan, Aug. 21-26, 1994

[5] S. G. Tantawi, et al. "Active radio frequency pulse compression using switched resonant delay lines" Nuclear Instruments & Methods in Physic Research, Section A (Accelerators, Spectrometers, Detectors and Associated Equipment) Elsevier, 21 Feb. 1996. Vol.370, No.2-3, pp. 297-302.

[6] Sami G. Tantawi et al: 'Active High-Power RF Pulse Compression Using Optically Switched Resonant Delay Lines', IEEE Trans. on Microwave Theory and Techniques, Vol. 45, No 8, pp. 1486, August, 1997

[7] Sami G. Tantawi, *et al.*, "The Generation of 400-MW RF Pulses at X Band Using Resonant Delay Lines," IEEE Trans. MTT, vol. 47, no. 12, December 1999; SLAC-PUB-8074.

[8] S.G. Tantawi, et al., "RF Components Using Over-Moded Rectangular Waveguides for the Next Linear Collider Multimoded Delay Line RF Distribution System," presented at the 18th Particle Accelerator Conference, New York, NY, March 29-April 2,1999.

[9] C.D. Nantista, *et al.*, "Planar Waveguide Hybrids for Very High Power RF," presented at the 1999 Particle Accelerator Conference, New York, NY, March 29-April 2, 1999; SLAC-PUB-8142.

[10] S.G. Tantawi, *et al.*, "A Multimoded RF Delay Line Distribution System for the Next Linear Collider," proc. of the Advanced Accelerator Concepts Workshop, Baltimore, MD, July 5-11, 1998, pp. 967-974.

[11] Sami G. Tantawi, et al., "Evaluation of the $TE_{12}$ Mode in Circular Waveguide for Low-Loss, High-Power RF Transmission," Phys. Rev. ST Accel. Beams, vol.3, 2000.

[12] Sami G. Tantawi and Mikhail I. Petelin: 'The Design and Analysis and Multi-Megawatt Distributed Single Pole Double Throw (SPDT) Microwave Switches', IEEE MTT-S Digest, p1153-1156, 1998

[13] Fumihiko Tamura and Sami G. Tantawi, "Multi-Megawatt X-Band Semiconductor Microwave Switches," IEEE MTT-S Digest, 1999

[14] S. G. Tantawi, et al. *"A Multimoded RF Delay Line Distribution System for the Next Linear Collider,"* Proce of the Advanced Accelerator Concepts Workshop, Baltimore, Maryland, July 5-11, 1998, p. 967-974

[15] P.B. Wilson, Z.D. Farkas, and R.D. Ruth, "SLED II: A New Method of RF Pulse Compression," Linear Accel. Conf., Albuquerque, NM, Sept. 1990; SLAC-PUB-5330.

[16] Z. H. Li et al, *"Mode Launcher Design for the Multimoded DLDS,"* Proc. of the 6th European Particle Accelerator Conference (EPAC 98), Stockholm, Sweden, 22-26 Jun 1998, p. 1900-1903.

[17] C. Nantista and Sami G. Tantawi, " A Planar Rectangular Waveguide Launcher and Extractor for a Dual-Moded RF Power Distribution System, " This proceedings.